\documentclass[12pt]{article}

\usepackage{epsfig,cite}

\catcode`@=11
\def\citer{\@ifnextchar
[{\@tempswatrue\@citexr}{\@tempswafalse\@citexr[]}}

\def\@citexr[#1]#2{\if@filesw\immediate\write\@auxout{\string\citation{#2}}\fi
  \def\@citea{}\@cite{\@for\@citeb:=#2\do
    {\@citea\def\@citea{--\penalty\@m}\@ifundefined
       {b@\@citeb}{{\bf ?}\@warning
       {Citation `\@citeb' on page \thepage \space undefined}}%
\hbox{\csname b@\@citeb\endcsname}}}{#1}}
\catcode`@=12

\begin{document}

%%%%%%%%%%%%%%%%%%%%%%%%%%%%%%%%%%%%%%%%%%%%%%%%%%%%%%%%%%%%%%%%%%%%%%

%%%%%%%%%%%%%%%%%%%%%%%%%%%%%% NEW SHORT CUTS %%%%%%%%%%%%%%%%%%%%%%%%

\input paperdef 

%%%%%%%%%%%%%%%%%%%%%%%%%%%%%% TITLE PAGE %%%%%%%%%%%%%%%%%%%%%%%%%%%%

\thispagestyle{empty}
\setcounter{page}{0}
\def\thefootnote{\fnsymbol{footnote}}

\mbox{}\vspace{-4em}
\begin{flushright}
ANL--HEP--CP--02--017 \hfill
BNL--HET--02/7\\
DCPT/02/24 \hfill
EFI--02--64\\
FERMILAB--Pub--02/024--T \hfill
IPPP/02/11\\
hep-ph/0202167 \\
\end{flushright}

\vspace{1em}

\begin{center}

{\large\sc {\bf Suggestions for Benchmark Scenarios}}

\vspace*{0.4cm} 

{\large\sc {\bf for MSSM Higgs Boson Searches at Hadron Colliders}}%
\footnote{Extended version of the contribution to the workshop
``Physics at TeV Colliders'', Les Houches, France,
\mbox{}\hspace{1.5em} May 2001}

\vspace{1cm}

{\sc M.~Carena$^{\,1}$%
\footnote{
email: carena@fnal.gov
}%
, S.~Heinemeyer$^{\,2}$%
\footnote{
email: Sven.Heinemeyer@physik.uni-muenchen.de
}%
, C.E.M.~Wagner$^{\,3,4}$%
\footnote{
email: cwagner@hep.anl.gov
}%
~and G.~Weiglein$^{\,5}$%
\footnote{
email: Georg.Weiglein@durham.ac.uk
}%
}

\vspace*{1cm}

$^1$ Theoretical Physics. Dept., Fermilab National Accelerator Lab.,\\
Batavia, IL 60510-0500, USA

\vspace*{0.4cm}

$^2$ HET, Physics Dept., Brookhaven Natl.\ Lab., Upton, NY 
11973, USA

\vspace*{0.4cm}

$^3$ HEP Division, Argonne Natl.\ Lab., 9700 Cass Ave.,
Argonne, IL 60439, USA

\vspace*{0.4cm}

$^4$ 
Enrico Fermi Institute, Univ.\ of Chicago, 5640 Ellis Ave.,
Chicago, IL 60637, USA

\vspace*{0.4cm}

$^5$ Institute for Particle Physics Phenomenology, University of Durham,\\
Durham DH1~3LR, UK

\end{center}

\vspace*{1cm}

\begin{abstract}

The Higgs boson search has shifted from LEP2 to the Tevatron and will
subsequently move to the LHC. Due to the different initial states, the
Higgs production and decay channels relevant for Higgs boson searches
were different at LEP2 to what they are at hadron colliders.
We suggest new benchmark scenarios for the MSSM Higgs boson search at
hadron colliders that exemplify the phenomenology of different parts of
the MSSM parameter space.
Besides the $\mhmax$ scenario and the no-mixing scenario 
used in the LEP2 Higgs boson searches, we propose two new
scenarios. In one the main production channel at the LHC, $gg \to h$,
is suppressed. In the other, important Higgs decay channels at
the Tevatron and at the LHC, $\hbb$ and $\htautau$, are suppressed. All
scenarios evade the LEP2 constraints for nearly the whole
$\MA$--$\tb$-plane. 

\end{abstract}

\def\thefootnote{\arabic{footnote}}
\setcounter{footnote}{0}

\newpage

%%%%%%%%%%%%%%%%%%%%%%%%%%%%%% LH TITLE PAGE %%%%%%%%%%%%%%%%%%%%%%%%%

\begin{center}
{\large\sc {\bf 
Suggestions for Benchmark Scenarios
for MSSM Higgs Boson Searches at Hadron Colliders
}}

\vspace{0.5cm}

{\sc
M. Carena, S. Heinemeyer, C.E.M. Wagner, G. Weiglein
}
\end{center}

\begin{abstract}
The Higgs boson search has shifted from LEP2 to the Tevatron and will
subsequently move to the LHC. Due to the different initial states, the
Higgs production and decay channels relevant for Higgs boson searches
were different at LEP2 to what they are at hadron colliders.
We suggest new benchmark scenarios for the MSSM Higgs boson search at
hadron colliders that exemplify the phenomenology of different parts of
the MSSM parameter space.
Besides the $\mhmax$ scenario and the no-mixing scenario 
used in the LEP2 Higgs boson searches, we propose two new
scenarios. In one the main production channel at the LHC, $gg \to h$,
is suppressed. In the other, important Higgs decay channels at
the Tevatron and at the LHC, $\hbb$ and $\htautau$, are suppressed. All
scenarios evade the LEP2 constraints for nearly the whole
$\MA$--$\tb$-plane. 
\end{abstract} 

%%%%%%%%%%%%%%%%%%%%%%%%%%%%%%%  MAIN TEXT  %%%%%%%%%%%%%%%%%%%%%%%%%%%%

\section{Introduction and theoretical basis}

Within the MSSM the masses of the $\cp$-even neutral Higgs bosons are
calculable in terms of the other MSSM parameters. The lightest Higgs
boson has been of particular interest, since its mass, $\mh$, is
bounded from above according to $\mh \leq \MZ$ at the tree level.
The radiative corrections at \onel\ order have 
been supplemented in the last years with the leading \twol\ corrections, 
performed by renormalization group (RG) 
methods~\cite{mhiggsRG1a}, by renormalization 
group improvement of the  one-loop effective potential 
calculation~\cite{mhiggsRG1b,mhiggsRG2a}, 
by two-loop effective  potential calculations~\cite{mhiggsEP},   
and in the Feynman-diagrammatic (FD)
approach~\cite{mhiggsletter,mhiggslong}. These calculations predict an
upper bound for $\mh$ of about $\mh \lsim 135 \gev$.%
\footnote{
This value holds for $\mt = 175 \gev$ and $\msusy = 1 \tev$. If $\mt$
is raised by $5 \gev$ then the $\mh$ limit is increased by about $5 \gev$; 
using $\msusy = 2 \tev$ increases the limit by about $2 \gev$.
}

After the termination of LEP, the Higgs boson search has now shifted to
the Tevatron and will later be continued at the LHC. Due to the large
number of 
free parameters, a complete scan of the MSSM parameter space is too
involved. Therefore at LEP the search has been performed in three
benchmark scenarios~\cite{benchmark}. Besides the $\mhmax$~scenario,
which has been used to obtain conservative bounds on
$\tb$~\cite{tbexcl}, and the no-mixing scenario, the
large-$\mu$~scenario had been designed to encourage the investigation
of flavor and decay-mode independent decay channels
(instead of focusing on the $\hbb$ channel). The investigation 
of these channels has lead to exclusion
bounds~\cite{mhLEP2001} that finally completely ruled out the
large-$\mu$~scenario.

The different environment at hadron colliders implies different Higgs
boson production channels and also different relevant decay channels
as compared to LEP. The main production modes at the Tevatron will be
$V^* \to V \phi$ ($V = W, Z, \phi = h, H, A$) 
and also $b\bar b \to b\bar b \phi$, while the
relevant decay modes will be $\phi \to b\bar b$ 
and $\phi \to \tau^+\tau^-$~\cite{runIIreport}.
At the LHC, on the other hand, the most relevant processes for a Higgs
boson with $\mh \le 135 \gev$ will be $gg \to h \to \ga\ga$ and
$t \bar t \to t \bar t h \to t \bar t b \bar b$.
Also the $W$~boson fusion channel, 
$WW \to h \to \tau^+\tau^-$ has been shown to be
promising~\cite{zeppitautau}. 
In order to investigate these different modes, we propose new
benchmark scenarios for the Higgs boson searches at hadron colliders.
Contrary to the new ``SPS''~benchmark scenarios proposed in
\citere{sps} for general SUSY
searches, the scenarios proposed here are designed specifically to
study the MSSM Higgs sector without assuming any particular soft
SUSY-breaking scenario and 
taking into account constraints only from the Higgs boson sector~itself.

The tree-level value for $\mh$ within the MSSM is determined by $\tb$, 
the $\cp$-odd Higgs-boson mass, $\MA$, and the $Z$-boson mass, $\MZ$. 
Beyond the tree-level, the main correction to $\mh$ stems from the 
$t$--$\Stop$-sector, and for large values of $\tb$ also from the 
$b$--$\Sbot$-sector (see \citere{benchmark} for our notations).
Accordingly, the most important parameters for the corrections to $\mh$
are $\mt$, $\msusy$ (in this work we assume that the soft
SUSY-breaking parameters for sfermions are equal:
$\msusy := \MstL = \MstR = \MsbL = \MsbR$), 
$\Xt$~($\equiv \At - \mu/\tb$), 
and $\Xb$~($\equiv \Ab - \mu\tb$); $A_{t,b}$ are
the trilinear Higgs sfermion couplings and $\mu$ is the Higgs mixing
parameter.
% The mass of the lightest $\cp$-even Higgs boson 
$\mh$ depends furthermore on the SU(2) gaugino mass
parameter, $M_2$ (the U(1) gaugino mass parameter is given by
$M_1 = 5/3\, \sw^2/\cw^2\, M_2$). 
At the two-loop level also the gluino mass, $\mgl$, enters the
prediction for $\mh$.

It should be noted in this context that the FD result has been obtained
in the on-shell (OS) renormalization scheme (the corresponding Fortran
code, that has been used for the analyses by the LEP collaborations, is
\fh~\cite{feynhiggs,feynhiggs2}), whereas the RG result has been 
calculated using the \msbar\ scheme; see \citere{bse} for details (the
corresponding Fortran code, also used by the LEP collaborations, is
\subh~\cite{mhiggsRG1a,mhiggsRG1b,bse}). 
While the corresponding shift in the parameter $\msusy$ turns out to be 
relatively small in general, sizable differences can occur between the 
numerical values of 
$\Xt$ in the two schemes; see \citeres{mhiggslong,bse}. For this reason
we specify below different values for $\Xt$ within the two approaches.

%%%%%%%%%%%%%%%%%%%%%%%%%%%%%%%%%%%%%%%%%%%%%%%%%%%%%%%%%%%%%%
%%%%%%%%%%%%%%%%%%%%%%%%%%%%%%%%%%%%%%%%%%%%%%%%%%%%%%%%%%%%%%

\section{The benchmark scenarios}

In this section we define four benchmark scenarios suitable for the
MSSM Higgs boson search at hadron colliders. 
In these scenarios the values of the $\Stop$~and $\Sbot$~sector as
well as the gaugino masses will be fixed, while $\tb$ and $\MA$ are
the parameters that are varied.
It has been checked that
the scenarios evade the LEP2 bounds~\cite{mhLEP2001} over a wide range
of the $\MA$-$\tb$-plane, where the variation should be chosen
according to
\BEA
&& 0.5 \le \tb \le 50, \quad \MA \le 1 \tev~.
\EEA

To illustrate the effects arising in the different proposed benchmark
scenarios, we have evaluated (using the results presented in 
\citeres{hdecay,deltamb1,deltamb2,hff})
\BE
\frac{\KKL \si \times \br \KKR_{\SU}}
     {\KKL \si \times \br \KKR_{\SM}}
\EE
for several Higgs production and decay channels for the new benchmark
scenarios. Further information about the properties of these scenarios
can be found at {\tt www.feynhiggs.de}~.

%%%%%%%%%%%%%%%%%%%%%%%%%%%%%%%%%%%%%%%%%%%%%%%%%%%%%%%%%%%%%%
%%%%%%%%%%%%%%%%%%%%%%%%%%%%%%%%%%%%%%%%%%%%%%%%%%%%%%%%%%%%%%

\subsection{The $\mhmax$ scenario}
\label{subsec:mhmax}

This scenario is kept as presented in \citere{benchmark},
since it allows conservative $\tb$ exclusion bounds~\cite{tbexcl} 
(only the sign of $\mu$ is switched to a positive value).
The parameters are chosen such that the maximum possible 
Higgs-boson mass as a function of $\tb$ is obtained
(for fixed $\msusy$, 
and $\MA$ set to its maximal value, $\MA = 1 \tev$).
The parameters are%
\footnote{
Better agreement with $\br(b \to s \ga)$ constraints is obtained for
the other sign of $\Xt$ (called the ``constrained $\mhmax$''
scenario). However, this lowers the maximum $\mh$ values by $\sim 5 \gev$.
}%
:
\BEA
&& \mt = 174.3 \gev, \quad \msusy = 1 \tev, \quad
\mu = 200 \gev, \quad M_2 = 200 \gev, \non \\
\label{mhmax}
&& \Xt^{\OS} = 2\, \msusy  \; \mbox{(FD calculation)}, \quad
   \Xt^{\MS} = \sqrt{6}\, \msusy \; \mbox{(RG calculation)} \non \\ 
&& \Ab = \At, \quad \mgl = 0.8\,\msusy~.
\EEA

%%%%%%%%%%%%%%%%%%%%% FIGURE %%%%%%%%%%%%%%%%%%%%%%%%%%%%%%%%%%%%%%%%%%
\begin{figure}[htb!]
\vspace{1em}
\begin{center}
\mbox{
\epsfig{figure=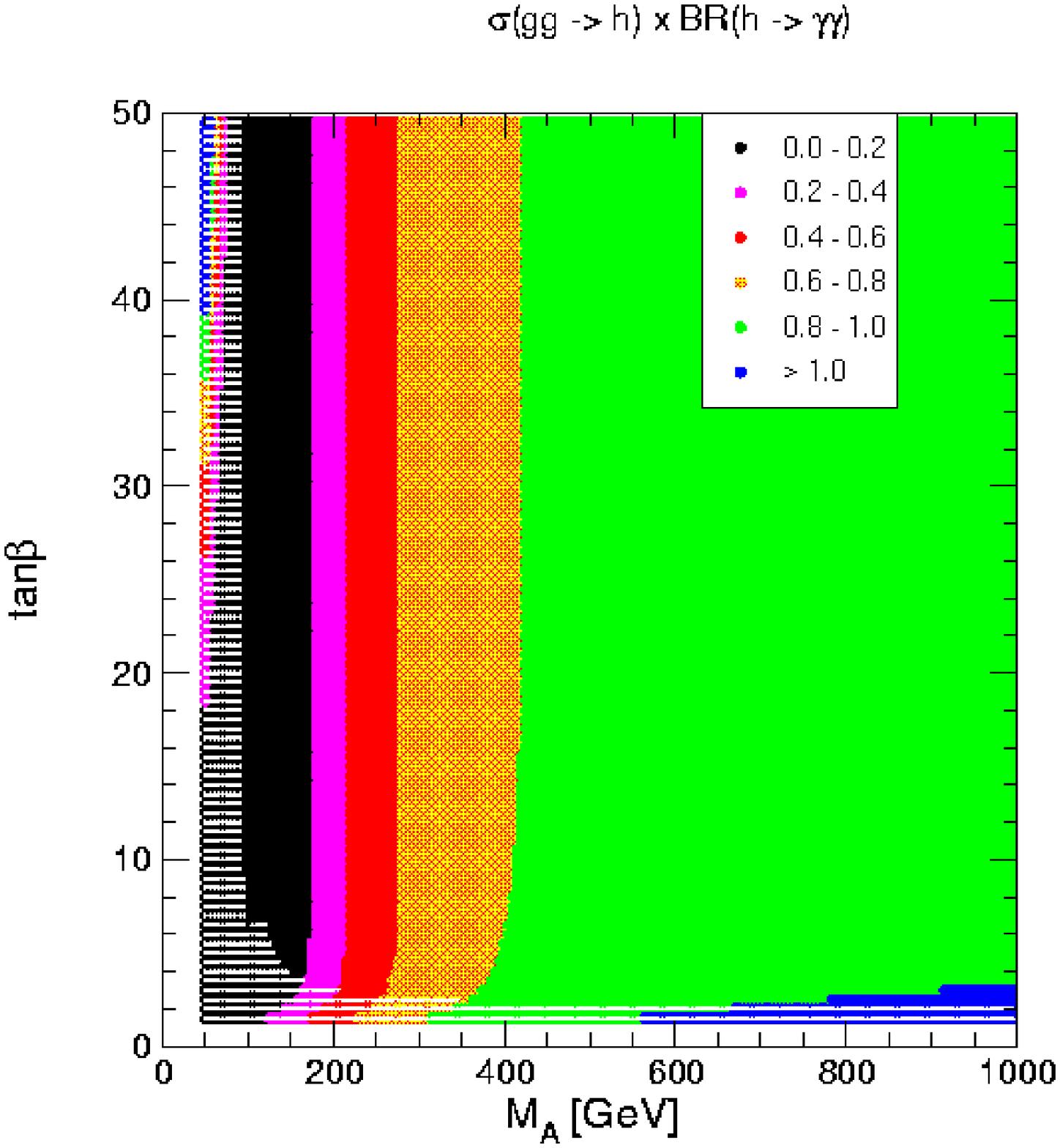,width=7cm,height=6.3cm}  
\hspace{1em}
\epsfig{figure=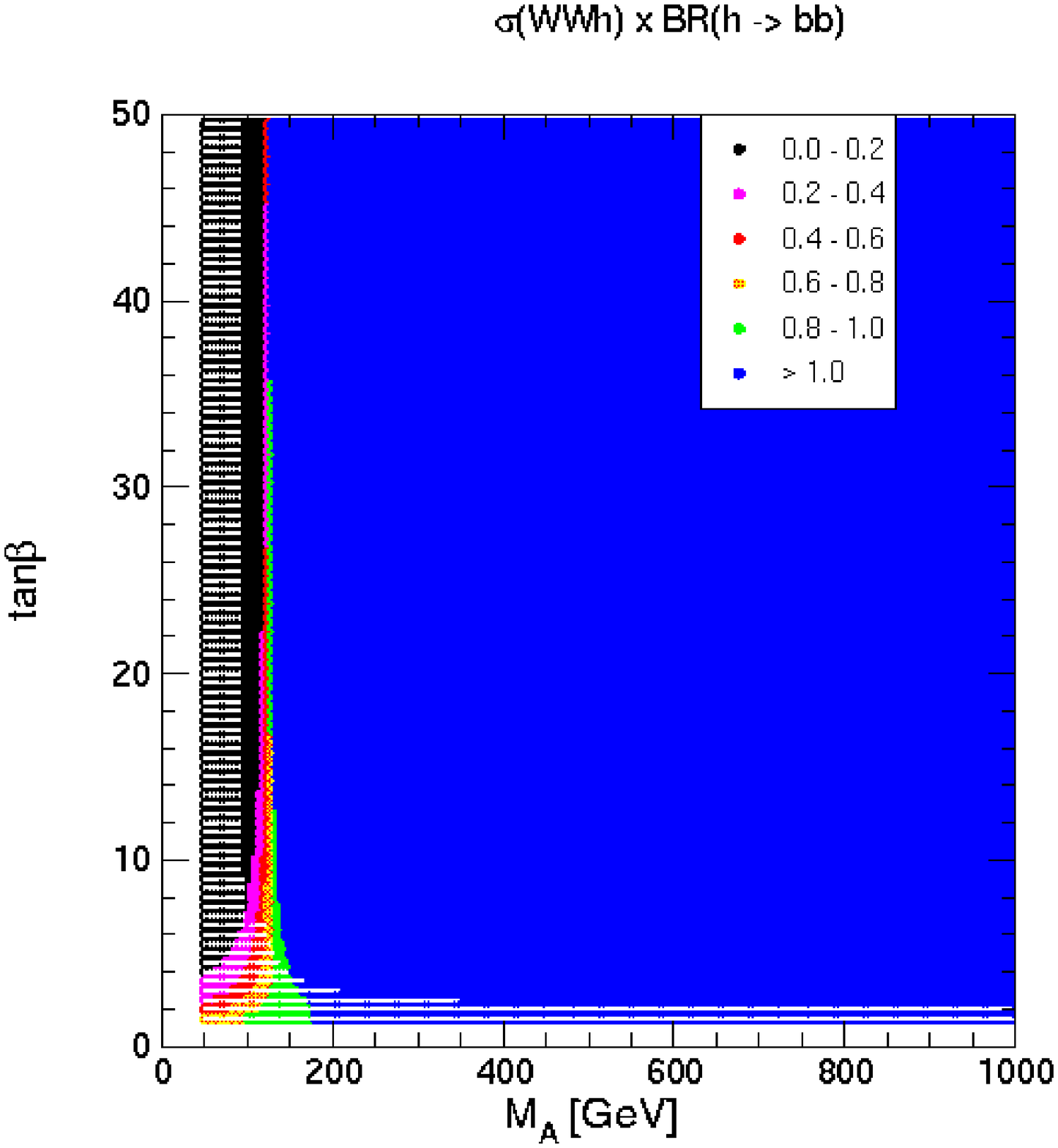,width=7cm,height=6.3cm} 
}
\end{center}
\caption[]{$[\si \times \br]_{\SU} / [\si \times \br]_{\SM}$ 
is shown for the channels $gg \to h \to \ga\ga$ (left plot) and 
$W^* \to W h \to W b\bar b$ (right plot) in the $\MA-\tb$-plane for the
$\mhmax$ scenario. The white-dotted area is excluded by LEP Higgs
searches. 
}
\label{fig:mhmax}
\end{figure}
%%%%%%%%%%%%%%%%%%%%% FIGURE %%%%%%%%%%%%%%%%%%%%%%%%%%%%%%%%%%%%%%%%%%

In \reffi{fig:mhmax} we show 
$[\si \times \br]_{\SU} / [\si \times \br]_{\SM}$ for the channels 
$gg \to h \to \ga\ga$ (left plot) and  
$W^* \to W h \to W b\bar b$ (right plot) in the $\MA-\tb$-plane. 
For low values of the $\cp$-odd Higgs boson mass, $\MA$, the 
structure of the MSSM Higgs sector leads to an enhancement with
respect to the SM value of the $hb\bar b$ and $h\tau^+\tau^-$
coupling. The MSSM coupling possesses an additional factor of 
$-\Saeff/\Cb$, where
$\aeff$ is the mixing angle of the neutral $\cp$-even Higgs sector,
including radiative corrections (see e.g. \citeres{deltamb2,hff}), and
the factor of $1/\Cb$ can lead to an enhancement for large $\tb$. 
The additional factor in the coupling converges very slowly to 1
for large values of $\MA$, and can differ 
significantly for low values of $\MA$. The branching ratio of the 
decay of $h$ into other channels, in particular the $h \to \ga\ga$
channel, is in this case significantly suppressed. Therefore, 
the $gg \to h \to \ga\ga$ channel can be strongly suppressed for small
and moderate values of $\MA$.
The $W^* \to W h \to W b\bar b$ channel, on the
other hand, is nearly always enhanced compared to the SM case, since
the $WWh$~coupling is mostly SM-like, except for $\MA \lsim 100\gev$, 
where for intermediate and large values of $\tb$ the $WWh$~coupling
becomes small. In this parameter region, however, 
for very small values of $\MA$ the heavy $\cp$-even Higgs boson,
$H$, has SM-like couplings. As a consequence, in the parameter regions
where the light Higgs boson is very difficult to observe, the search
for the $H$ should have similar prospects as the SM Higgs search.

The channels presented in \reffi{fig:mhmax} show the complementarity
between Tevatron and LHC for the search for the lightest MSSM Higgs boson.
It is worth to notice as well that the LHC itself will exhibit a similar
complementary behavior in this benchmark scenario between the
$gg \to h \to \ga\ga$ process and the other two search channels, 
$t\bar t h \to t\bar t b\bar b$ and $WW \to h \to \tau^+\tau^-$.

%%%%%%%%%%%%%%%%%%%%%%%%%%%%%%%%%%%%%%%%%%%%%%%%%%%%%%%%%%%%%%
%%%%%%%%%%%%%%%%%%%%%%%%%%%%%%%%%%%%%%%%%%%%%%%%%%%%%%%%%%%%%%

\subsection{The no-mixing scenario}

This benchmark scenario is the same as the $\mhmax$ scenario, but with
vanishing mixing in the $\Stop$~sector and with a higher SUSY mass
scale to avoid the LEP Higgs bounds:
\BEA
&& \mt = 174.3 \gev, \quad \msusy = 2 \tev, \quad
\mu = 200 \gev, \quad M_2 = 200 \gev, \non \\
\label{nomix}
&& \Xt = 0  \; \mbox{(FD/RG calculation)}, \quad
   \Ab = \At, \quad \mgl = 0.8\,\msusy~.
\EEA

%%%%%%%%%%%%%%%%%%%%% FIGURE %%%%%%%%%%%%%%%%%%%%%%%%%%%%%%%%%%%%%%%%%%
\begin{figure}[htb!]
\vspace{1em}
\begin{center}
\mbox{
\epsfig{figure=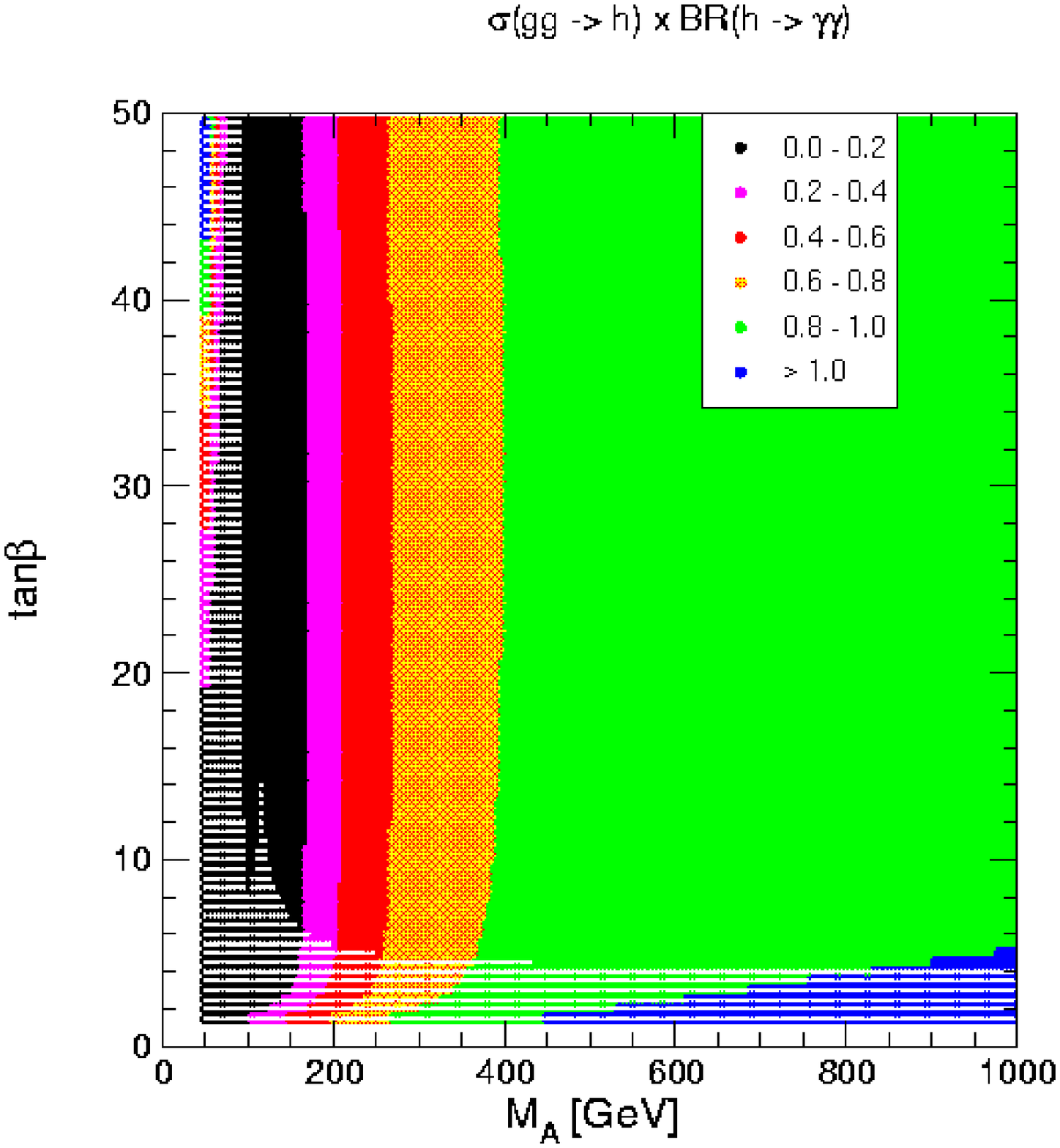,width=7cm,height=6.3cm}  
\hspace{1em}
\epsfig{figure=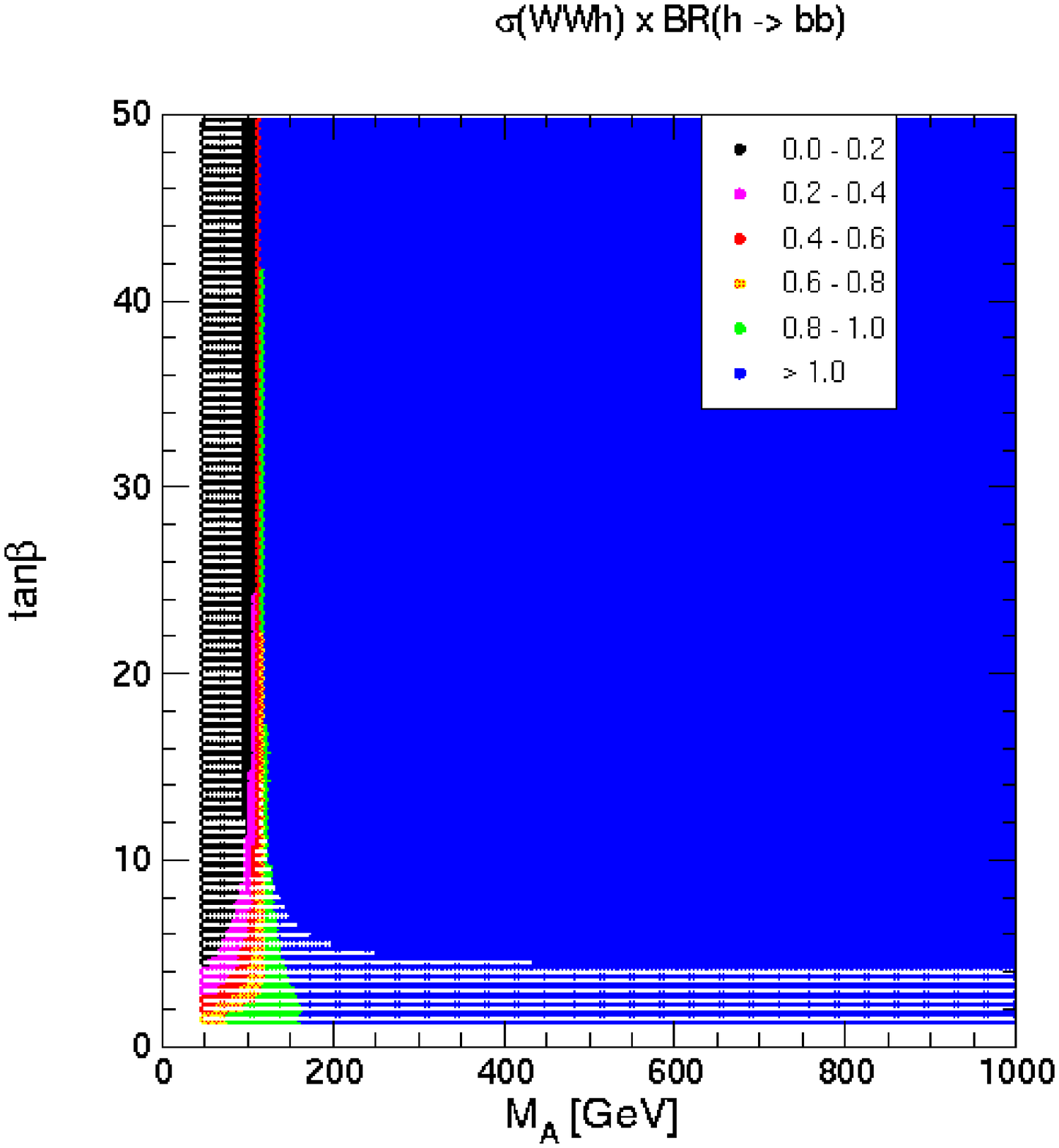,width=7cm,height=6.3cm} 
}
\end{center}
\caption[]{$[\si \times \br]_{\SU} / [\si \times \br]_{\SM}$ 
is shown for the channels $gg \to h \to \ga\ga$ (left plot) and 
$W^* \to W h \to W b\bar b$ (right plot) in the $\MA-\tb$-plane for the
no-mixing scenario. The white-dotted area is excluded by LEP Higgs
searches. 
}
\label{fig:nomix}
\end{figure}
%%%%%%%%%%%%%%%%%%%%% FIGURE %%%%%%%%%%%%%%%%%%%%%%%%%%%%%%%%%%%%%%%%%%

In \reffi{fig:nomix} we show 
$[\si \times \br]_{\SU} / [\si \times \br]_{\SM}$ for the channels 
$gg \to h \to \ga\ga$ (left plot) and  
$W^* \to W h \to W b\bar b$ (right plot) in the $\MA-\tb$-plane. As in the
$\mhmax$~scenario, the branching ratio for $\hbb$ and $\htautau$
is significantly enhanced for low values of $\MA$. Therefore, also in
the no-mixing scenario the  
$gg \to h \to \ga\ga$ channel can be strongly suppressed for not too
large values of $\MA$. The $W^* \to W h \to W b\bar b$ channel
is nearly always enhanced compared to the SM case for the same reasons
as in the $\mhmax$~scenario (except for $\MA \lsim 100 \gev$).

%%%%%%%%%%%%%%%%%%%%%%%%%%%%%%%%%%%%%%%%%%%%%%%%%%%%%%%%%%%%%%
%%%%%%%%%%%%%%%%%%%%%%%%%%%%%%%%%%%%%%%%%%%%%%%%%%%%%%%%%%%%%%

\subsection{The gluophobic Higgs scenario}

In this scenario the main production cross section for the light Higgs
boson at 
the LHC, $gg \to h$, is strongly suppressed. This can happen due to a
cancellation between the top quark and the stop quark loops in the
production vertex (see \citere{ggsuppr}). This cancellation is more
effective for small $\Stop$~masses and hence for relatively large
values of the $\Stop$~mixing parameter, $\Xt$. The partial width of
the most relevant decay mode, $\Ga(h \to \ga\ga)$, is affected much
less, since it is dominated by the $W$~boson loop.
The parameters are:
\BEA
&& \mt = 174.3 \gev, \quad \msusy = 350 \gev, \quad
\mu = 300 \gev, \quad M_2 = 300 \gev, \non \\
\label{ggsup}
&& \Xt^{\OS} = -750 \gev \; \mbox{(FD calculation)}, \quad
   \Xt^{\MS} = -770 \gev \; \mbox{(RG calculation)} \non \\ 
&& \Ab = \At, \quad \mgl = 500 \gev~.
\EEA

%%%%%%%%%%%%%%%%%%%%% FIGURE %%%%%%%%%%%%%%%%%%%%%%%%%%%%%%%%%%%%%%%%%%
\begin{figure}[htb!]
\vspace{1em}
\begin{center}
\mbox{
\epsfig{figure=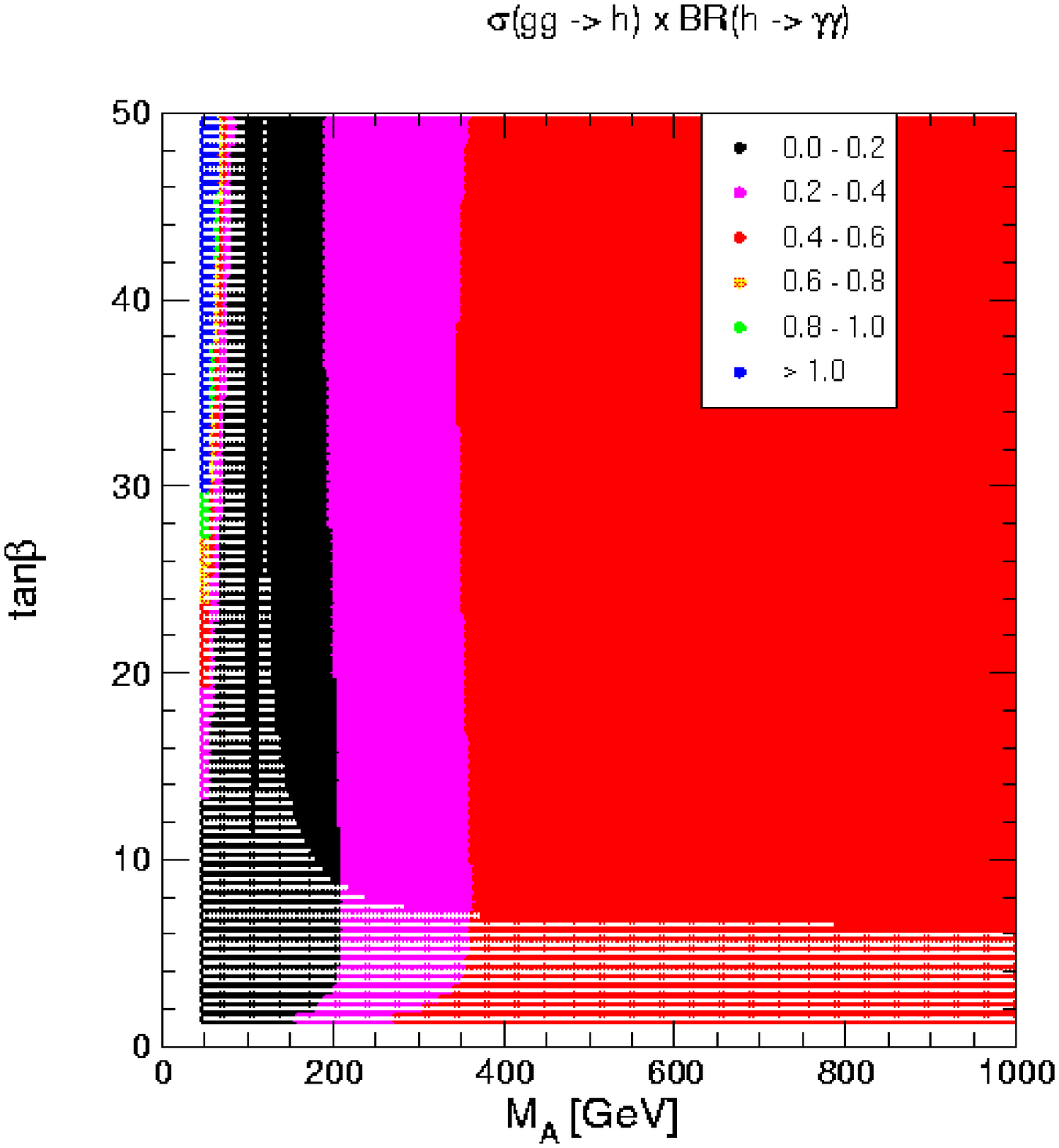,width=7cm,height=6.3cm}  
\hspace{1em}
\epsfig{figure=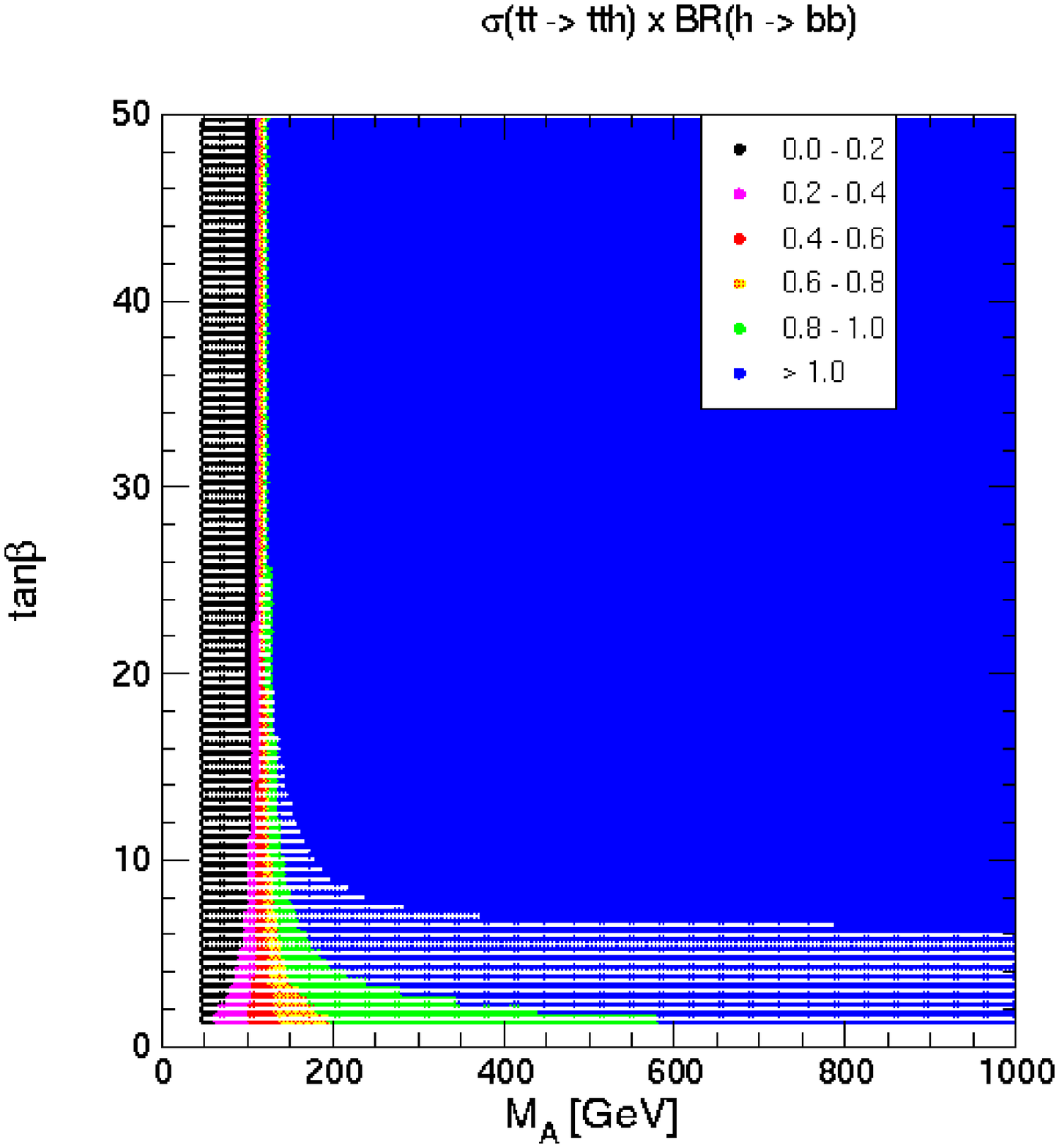,width=7cm,height=6.3cm} 
}
\end{center}
\caption[]{$[\si \times \br]_{\SU} / [\si \times \br]_{\SM}$ 
is shown for the channels $gg \to h \to \ga\ga$ (left plot) and 
$t\bar t \to t\bar t h \to t \bar t b\bar b$ (right plot) 
in the $\MA-\tb$-plane for the
gluophobic Higgs scenario. The white-dotted area is excluded by LEP Higgs
searches. 
}
\label{fig:gluophobicHiggs}
\end{figure}
%%%%%%%%%%%%%%%%%%%%% FIGURE %%%%%%%%%%%%%%%%%%%%%%%%%%%%%%%%%%%%%%%%%%

In \reffi{fig:gluophobicHiggs} we show 
$[\si \times \br]_{\SU} / [\si \times \br]_{\SM}$ for the channels 
$gg \to h \to \ga\ga$ (left plot) and  
$t\bar t \to t\bar t h \to t \bar t b\bar b$ (right plot) 
in the $\MA-\tb$-plane. 
The $gg \to h \to \ga\ga$ channel can be strongly suppressed over the
whole $\MA-\tb$-plane, rendering this detection channel difficult.
The $t\bar t \to t\bar t h \to t \bar t b\bar b$ channel, on the other
hand, is always enhanced compared to the SM case (except for 
$\MA \lsim 100 \gev$). The same qualitative behavior holds for the
$WW$~fusion channel with subsequent decay to $b\bar b$ or 
$\tau^+\tau^-$.

%%%%%%%%%%%%%%%%%%%%%%%%%%%%%%%%%%%%%%%%%%%%%%%%%%%%%%%%%%%%%%
%%%%%%%%%%%%%%%%%%%%%%%%%%%%%%%%%%%%%%%%%%%%%%%%%%%%%%%%%%%%%%

\subsection{The small $\aeff$ scenario}

Besides the channel $gg \to h \to \ga\ga$ at the LHC, the other
channels for light Higgs 
searches at the Tevatron and at the LHC mostly rely on the decays $\hbb$
and $\htautau$, see \refse{subsec:mhmax}.
If $\aeff$ is 
small, these two decay channels can be heavily suppressed in the MSSM
due to the additional factor $-\Saeff/\Cb$ compared to the SM coupling. 
($\hbb$ can also receive large corrections from
$\Sbot$-$\gl$~loops~\cite{deltamb1,deltamb2}.) 
Such a suppression occurs for large $\tb$ and not too large $\MA$ (in a
similar way as in the large-$\mu$ scenario~\cite{benchmark}) for the
following parameters: 
\BEA
&& \mt = 174.3 \gev, \quad \msusy = 800 \gev, \quad
\mu = 2.5 \, \msusy, \quad M_2 = 500 \gev, \non \\
\label{smallaeff}
&& \Xt^{\OS} = -1100 \gev \; \mbox{(FD calculation)}, \quad 
   X_t^{\MS} = -1200 \gev \; \mbox{(RG calculation)} \non \\ 
&& \Ab = \At, \quad \mgl = 500 \gev~.
\EEA

%%%%%%%%%%%%%%%%%%%%% FIGURE %%%%%%%%%%%%%%%%%%%%%%%%%%%%%%%%%%%%%%%%%%
\begin{figure}[htb!]
\vspace{1em}
\begin{center}
\mbox{
\epsfig{figure=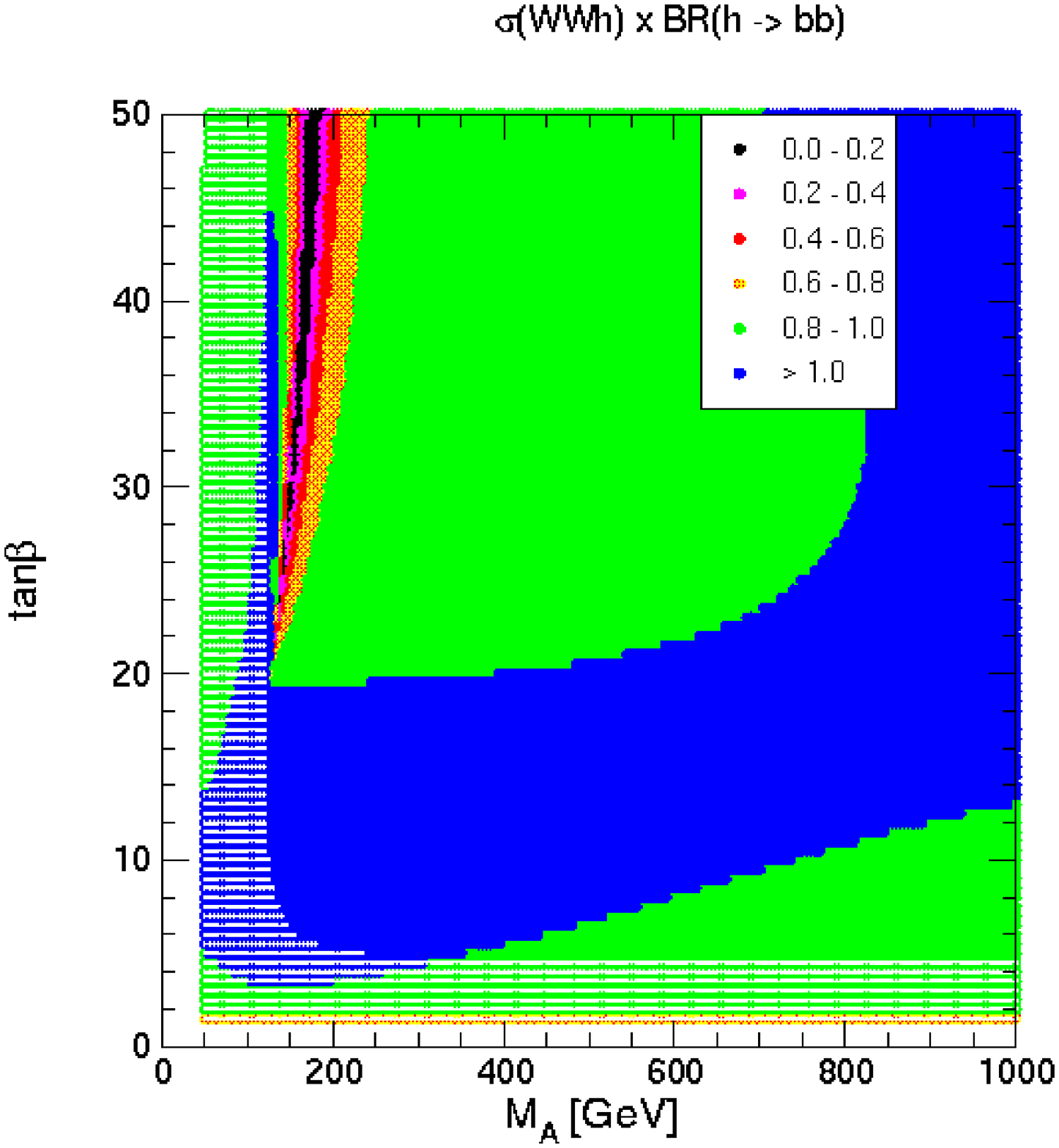,width=7cm,height=6.3cm} 
\hspace{1em}
\epsfig{figure=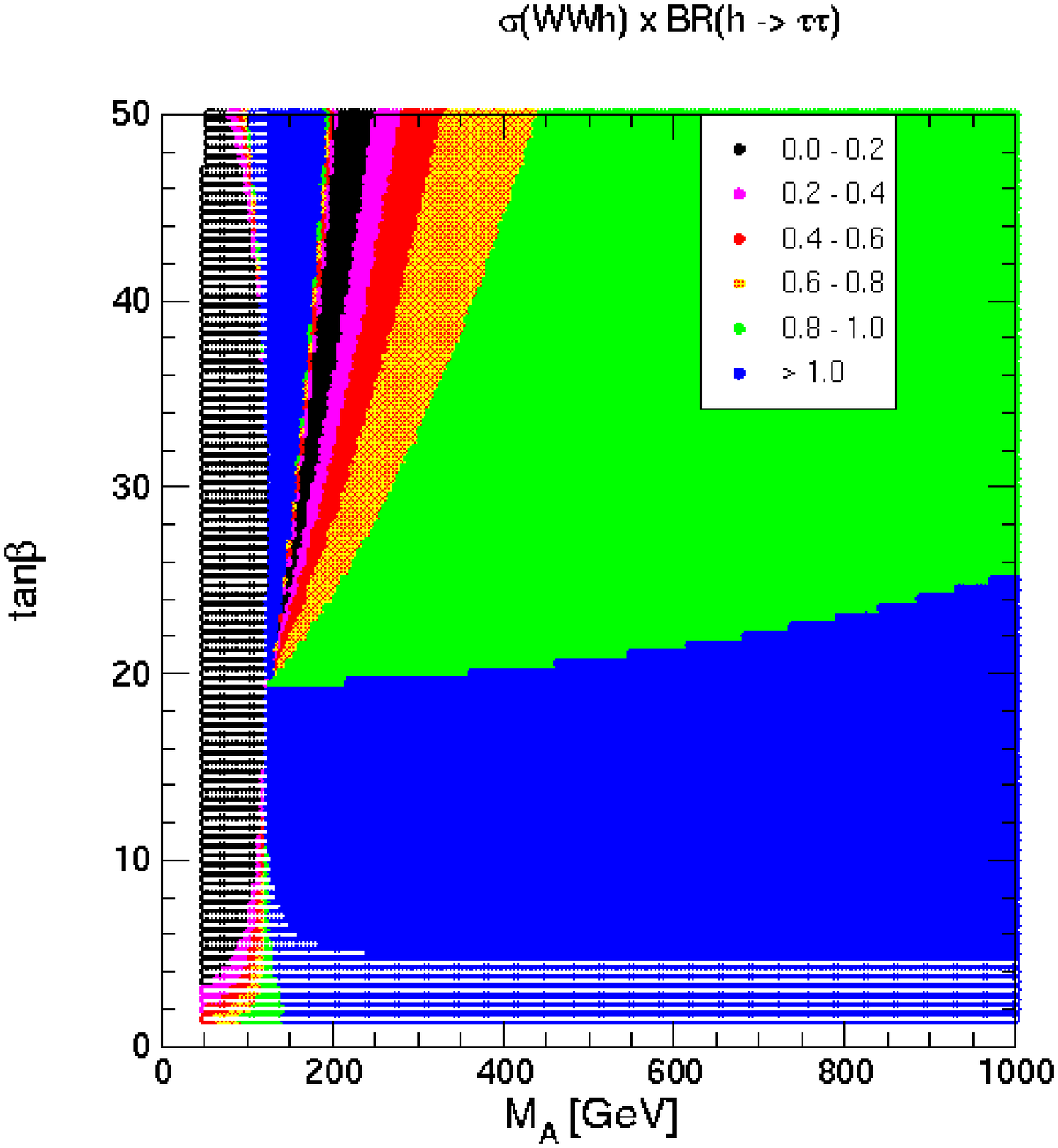,width=7cm,height=6.3cm}  
}
\end{center}
\caption[]{$[\si \times \br]_{\SU} / [\si \times \br]_{\SM}$ 
is shown for the channels 
$W^* \to W h \to W b\bar b$ (left plot) and
$W^* \to W h \to W \tau^+\tau^-$ (right plot) 
in the $\MA-\tb$-plane for the
small $\aeff$~scenario. The white-dotted area is excluded by LEP Higgs
searches. 
}
\label{fig:smallaleff}
\end{figure}
%%%%%%%%%%%%%%%%%%%%% FIGURE %%%%%%%%%%%%%%%%%%%%%%%%%%%%%%%%%%%%%%%%%%

In \reffi{fig:smallaleff} we show 
$[\si \times \br]_{\SU} / [\si \times \br]_{\SM}$ for the channels 
$W^* \to W h \to W b\bar b$ (left plot) and
$W^* \to W h \to W \tau^+\tau^-$ (right plot) 
in the $\MA-\tb$-plane. 
Significant suppression occurs for large $\tb$, $\tb \gsim 20$, and
small $\MA$, $\MA \lsim 250, 400 \gev$, for $\hbb$ and $\htautau$, 
respectively. Thus, Higgs boson search via the $W$~production channel
and the $WW$~fusion channel (see also \citere{zeppitautau}) will be
difficult in these  parts of the parameter space. The same applies for
the channel $t\bar t h \to t\bar t b\bar b$. The complementary
channel, $h \to \ga\ga$, is unsuppressed compared to the SM case for
large parts of the $\MA-\tb$-plane.

%%%%%%%%%%%%%%%%%%%%%%%%%%%%%%%%%%%%%%%%%%%%%%%%%%%%%%%%%%%%%%%%%%%%%%%%
%%%%%%%%%%%%%%%%%%%%%%%%%%%%%%%%%%%%%%%%%%%%%%%%%%%%%%%%%%%%%%%%%%%%%%%%

\section{Conclusions}

We have presented four benchmark scenarios for the MSSM Higgs boson
search at hadron colliders, evading the exclusion bounds obtained at
LEP2. The different scenarios exemplify 
different features of the MSSM parameter space, such as large $\mh$
values and significant $gg \to h$ or $\hbb$, $\htautau$
suppression. 
In the benchmark scenarios proposed above, we have briefly analyzed 
the possible suppression of several Higgs production and decay
channels, showing possible ``problematic'' regions of the MSSM
parameter space. 

With the exception of the gluon fusion  mediated process, which is
significantly suppressed in the gluophobic Higgs scenario, the
production processes at the Tevatron and the LHC considered here, 
$W^* \to Wh$, $t\bar t \to t\bar t h$, $WW \to h$ and $gg \to h$, 
are close to their SM values for most of the allowed parameter space
of the benchmark scenarios.
Hence, deviations of the rate of the Higgs search processes at hadron
colliders compared to the SM case are mainly due to the SUSY
corrections affecting the Higgs decay modes. 

In all the cases analyzed in this
note, we have found a complementarity between the $\hbb$,
$\htautau$ and the $h \to \ga\ga$ channels, i.e.\ in the parameter
regions where the search for the Higgs in one of these channels becomes
problematic, in at least one of the other channels it becomes easier
than in the SM case. It is difficult to find exceptions to this rule in
the MSSM parameter space. Therefore 
in analyzing the new benchmark scenarios, it will be helpful to make use
of the complementarity of different channels accessible at the
Tevatron and the LHC (see e.g.\ \citere{deltamb2} for details).

%%%%%%%%%%%%%%%%%%%%%%%%%%%%%%%%%%%%%%%%%%%%%%%%%%%%%%%%%%%%%%%%%%%%%%%%
%%%%%%%%%%%%%%%%%%%%%%%%%%%%%%%%%%%%%%%%%%%%%%%%%%%%%%%%%%%%%%%%%%%%%%%%

\subsubsection*{Acknowledgements}

G.W. thanks the organizers of the Les Houches workshop for the
invitation and the pleasant and constructive atmosphere. 
%Only the opening hours of the bar could have been longer.

%%%%%%%%%%%%%%%%%%%%%%%  REFERENCES  %%%%%%%%%%%%%%%%%%%%%%%

\bibliographystyle{plain}

\end{document}